# Physicochemical and thermal modelling of the reaction between a porous pellet and a gas

**F. Patisson, D. Ablitzer**

Laboratoire de Science et Génie des Matériaux et de Métallurgie,

UMR 7584 CNRS-INPL,

École des Mines,

Parc de Saurupt,

54042 Nancy Cedex,

France.

**Abstract**

A mathematical model has been developed to simulate the kinetic and thermal behaviour of a porous solid pellet undergoing chemical reaction with a gas. The model describes the chemical reaction itself, the transfer of the gaseous species between the external gas and the pellet surface, the transport of these species to the inside of the pellet through the pores and through the layer of solid reaction products, the generation (or consumption) of heat due to the reaction and the associated heat transfer processes, together with the structural changes produced in the solid by the reaction. The model has been validated by comparison with experimental results and data from the literature. Simulation results are presented for two reactions: the exothermal oxidation of zinc sulphide and the hydrofluorination of uranium dioxide.

**Key words**

Gas-solid reaction; mathematical model; kinetics; diffusion; heat transfer; porosity.

# 1. Introduction

Numerous industrial processes involving the conversion of divided solids (e.g. ore treatment by roasting and reduction, coal pyrolysis, the combustion of solids, waste incineration, the absorption of sour gases by lime) are controlled by the kinetics of gas-solid reactions of the type

$$a A_{(g)} + b B_{(s)} \rightleftarrows p P_{(g)} + q Q_{(s)} \qquad (1)$$

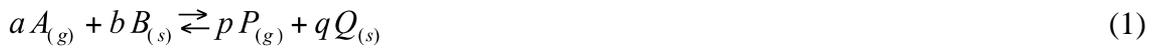

where $a$, $b$, $p$ and $q$ are the stœchiometry coefficients, $A$ and $B$ the gaseous and solid reactants, and $P$ and $Q$ are the reaction products.

The porous solid particles treated industrially are generally pellets made up of a number of grains, of different sizes, separated by pores. If these grains are themselves porous, they can be considered as a collection of dense crystallites and pores. A possible representation of the solid is then that shown in Fig. 1a, where the pellet is in fact an agglomerate of dense grains of different sizes. Let us consider that the reaction (1) is taking place inside a particle of this sort immersed in a gas, and examine the associated physicochemical and thermal phenomena in more detail.

*1.1. The phenomena involved*

Before reacting with the solid $B$, the reactive gas $A$ must be transported to the reaction site by *(i)* transfer within the external gas to the surface of the pellet and *(ii)* diffusion through the intergranular pores. On the scale of each grain, the reaction can itself be broken down into several steps - adsorption *(iii-a)*, reaction at the external gas/$Q$ interface *(iii-b)*, reaction at the internal $B/Q$ interface *(v)*, desorption *(iii-c)* - and generally involves a final transport step *(iv)* through the layer of solid reaction product $Q$, which may be dense or porous (Fig. 1b). The gas produced $P$ is desorbed, then evacuated outwards through the pores. Each of these steps has its own kinetics and can limit, or help to limit, the overall rate of conversion. The situation will thus be described either as a chemical regime, when a surface process controls the overall kinetics, a diffusional regime when the rate is determined by diffusion, either in the gas inside the pores or in the layer of reaction product $Q$, as an external transfer limitedregime, or as a mixed regime. The presence of two interfaces, gas/$Q$ and $B/Q$, together with reaction/transport coupling are characteristic features of heterogeneous reactions of type (1).

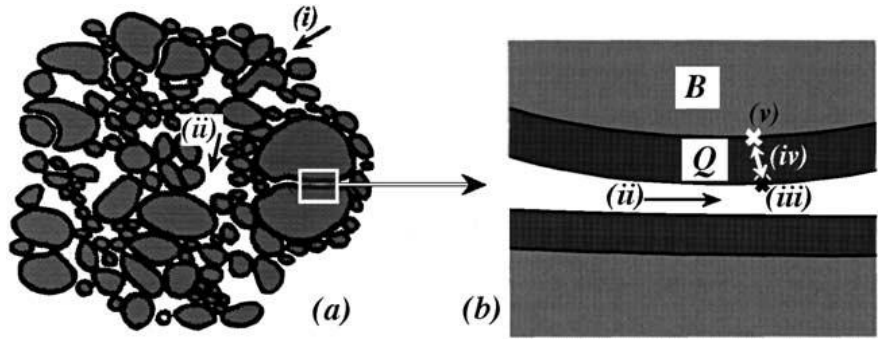

Fig. 1. Schematic representation of a porous pellet (a), a pore (b), and the steps (i–v) in mass transport.

If the reaction is exo- or endothermic, or if the external temperature varies, the associated heat evolution or consumption phenomena, and heat transfer both within the pellet and with the exterior are superimposed on the matter transport processes.

Finally, a characteristic of numerous gas-solid reactions is the change in the porous structure of the solid as the reaction proceeds. This change can be caused by external stresses, by high temperature sintering, or, as in the case of interest here, by the chemical reaction itself. The transformation of the solid phase $B$ to a new solid phase $Q$, of different molar volume, modifies the morphology of the solid on the scale of the grains, the pores, and sometimes the whole pellet. If the stœchiometric ratio between the molar volumes

$$Z = \frac{q\,v_{m,Q}}{b\,v_{m,B}} \qquad (2)$$

is greater than 1, as in the case of the reaction studied in § 3.2, the grains swell and the porosity decreases. If it is less than 1, as in the case of gasification reactions, for example, the porosity increases.

*1.2. Gas-solid reaction models*

Because of the economic importance of the solid conversion processes mentioned and the diversity of the reactions involved, a large number of models describing the reaction kinetics have been published. The aims of these models are to help to interpret the experimentally observed kinetics, to indicate the rate-controlling mechanisms, and to predict conversion rates under different processing conditions. They are sometimes even designed to be incorporated in reactor or process models.

Schematically, the different models can be divided into three categories: the homogeneous models, which consider the porous solid as a continuum [1], the grain models, which describe the solid phase as a juxtaposition of dense objects (the grains) [2-4], and the pore models, which represent the porous medium as a collection of hollow objects (the pores) [5-7]. Only a few of these models are mentioned here, moving directly from the original models to the most recent work. More detailed bibliographic studies can be found elsewhere [1,3].

Even though they belong to different categories in the above classification, these models are not really completely independent. They all seek to describe the same physical phenomena, with an increasing degree of fineness and sophistication, and the structural description in terms of either grains or pores essentially leads to differences in the calculation of the area of the reaction surface.

A particular description of the porous structure is therefore not the only criterion on which to base the choice of a model. Phenomena with an essential influence on the reaction kinetics of industrial divided solids processing, such as diffusional limitations of matter transport or variations in temperature, are often ignored. A particular feature of the model presented here is to simultaneously describe the kinetics, the thermal aspects and the variations in porosity.

## 2. Mathematical model

*2.1. Principle and assumptions*

The present model describes the kinetic and thermal behaviour of a porous pellet in which a gas-solid reaction of type (1) is occurring. It allows for the external transfer of the gaseous species to the pellet surface, diffusional transport in the gas inside the pores, the heterogeneous reaction, the generation or consumption of heat by the reaction, heat transfer by effective conduction and heat exchanges with the outside. It is a model of the homogeneous type, to which a pore model has been added to describe the porous structure and its evolution.

The reaction can be reversible, non-equimolar ($a \neq p$), and exo- or endothermic. The possible presence of inert gaseous ($I$) or solid ($J$) species is taken into account. The regime is transient and not quasi-stationary. Different reaction rate laws can be considered, and different diffusion regimes are possible. It is assumed that the pellet is spherical with a constant diameter and that the total gas pressure remains constant.

The model is of the unidimensional transient regime type. The variables and parameters are therefore functions of $r$, the radial position within the pellet, and of the time $t$.

## 2.2. Equations

Conservation equations

The total molar balance and those for $A$ and $P$ can be written:

$$\nabla \cdot \mathbf{N}_A + \frac{\partial}{\partial t}\left(\varepsilon c_t x_A\right) = -a r_V \tag{3}$$

$$\nabla \cdot \mathbf{N}_P + \frac{\partial}{\partial t}\left(\varepsilon c_t x_P\right) = p r_V \tag{4}$$

$$\nabla \cdot \mathbf{N}_t + \frac{\partial}{\partial t}\left(\varepsilon c_t\right) = (p - a) r_V \tag{5}$$

where $N_A$, $N_P$ and $N_t$ are the molar flux densities, $\varepsilon$ is the porosity, $c_t$ is the total concentration, $x_A$ and $x_P$ are the molar fractions of $A$ and $P$, and $r_V$ is the rate of reaction per unit volume of the pellet. The solid molar balances for $B$, $Q$ and $J$ are given by:

$$c_{B_0} \frac{\partial x}{\partial t} = b r_V \tag{6}$$

$$\frac{\partial c_Q}{\partial t} = q r_V \tag{7}$$

$$c_J = c_{J_0} \tag{8}$$

where $c_{B_0}$ and $c_{J_0}$ are the apparent initial molar concentrations of B and J, x is the local degree of conversion, and $c_Q$ and $c_J$ are the apparent molar concentrations of Q and J. Finally, the heat balance assumes that heat transport and storage in the gases are negligible and that the gas and solid are at the same temperature at all points in the pellet:

$$c_{pV}\frac{\partial T}{\partial t} + \nabla \cdot \left(-\lambda_e \nabla T\right) = r_V \left(-\Delta_r H\right) \tag{9}$$

where $c_{pV}$ is the specific heat per unit volume at constant pressure, $\lambda_e$ is the effective thermal conductivity and $\Delta_r H$ is the heat of reaction. Neglecting the heat transport by the gases in the thermal balance is a reasonable assumption at usual low total flux density $N_t$ (*e.g.* counter-diffusion of A and P). It may not apply to reactions (*e.g.* pyrolysis) generating a large net gas flow.

Expression of the molar flux densities

To calculate the diffusion fluxes, it is necessary to consider both the multi-component nature of the gas and its transport in the pores of a solid (possibility of a Knudsen regime, influence of pore size). In the present case, a simplified formulation of the fluxes is used, involving the total flux and effective diffusion coefficients for each species. Thus,

$$\mathbf{N}_A = x_A \mathbf{N}_t - D_{A_e} c_t \nabla x_A \tag{10}$$

$$\mathbf{N}_P = x_P \mathbf{N}_t - D_{P_e} c_t \nabla x_P \tag{11}$$

The effective diffusion coefficients $D_{A_e}$ and $D_{P_e}$ are functions of the composition of the mixture, of the binary diffusivities and of the porous structure. Different diffusion models, adapted to each problem, can be used to obtain these functions [1].

Expression of the reaction rate

The rate of a heterogeneous reaction $r_s$, defined per unit area of reaction surface, is usually written

$$r_s = k_r \left( c_{A_s}^n - \frac{c_{P_s}^l}{K_{eq}} \right) \quad \text{with} \quad k_r = k_0 e^{-\frac{E_a}{RT}} \tag{12}$$

where $k_r$ is the rate constant, assumed to obey Arrhenius' law, $c_{A_s}$ and $c_{P_s}$ are the molar concentrations at the reaction surface, n and l are the partial orders of reaction with respect to *A* and *P*, and $K_{eq}$ is the equilibrium constant. Other kinetic laws, such as that given by the Langmuir-Hinshelwood model, are also provided for in the present model.

If the chemical reaction is limited by a surface process, then the reaction rate per unit volume $r_V$ involved in the balance equations is directly related to the reaction rate per unit surface area via the specific area of the reaction surface. Thus

$$r_V = r_s a_0 s(x) \tag{13}$$

where $a_0$ is the initial specific area and where the function $s(x)$ reflects the change in the area of the reaction surface as a function of the local degree of conversion. This function depends on the structural model chosen and on the location of the rate-controlling surface chemical process. For example,

$$s(x) = (1-x)^m \quad \text{with} \quad m = \tfrac{2}{3} \tag{14}$$

corresponds to a reaction at the *B/Q* interface according to the shrinking-core spherical grain model [8]. An alternative is presented below.

*2.3. Pore sub-model*

In order to obtain a finer description of the solid when it contains a significant distribution of pore sizes and to simulate the reactions leading to a change in the porous structure, Bhatia's randomly distributed pore model [6] is used. The initial porous structure (pore diameters, porosity, specific area) is completely described with the aid of a unique function $l(R_0)$ such that $l(R_0)dR_0$ represents the total length of pores of initial radius $R_0$ per unit volume of pellet. This function can be simply derived from mercury porometry measurements.

Progress of the reaction leads to the displacement of the gas/*Q* and *B/Q* interfaces, characterized by their radii $R_p(R_0,t)$ and $R_r(R_0,t)$ (Fig. 2). The latter are determined from the following equations

$$\frac{\partial R_r}{\partial t} = \frac{b v_{mB} k_r c_A}{1 + \dfrac{a k_r (1-x)(1-e_0) R_r y}{D_{A(Q)}}} \tag{15}$$

$$\frac{\partial R_p}{\partial t} = -\frac{(Z-1)(1-x)}{1+(Z-1)(1-x)} \frac{R_r}{R_p} \frac{\partial R_r}{\partial t} \tag{16}$$

$$\frac{\partial y}{\partial t} = \frac{1}{1-\varepsilon_0}\left[\frac{1}{R_r(1-x)} + \frac{(Z-1)(1-x)R_r}{R_p^2(1+(Z-1)x)^2}\right]\frac{\partial R_r}{\partial t} \tag{17}$$

where $y$ is an intermediate calculation variable and $D_{A(Q)}$ is the diffusion coefficient for $A$ through the layer of reaction product $Q$. These expressions assume that the reaction is irreversible, of first order, and situated at the internal $B/Q$ interface, so that the use of this sub-model is restricted to reactions of this type. Knowing $R_r$ and $R_p$, it is possible to determine at each instant and each point the specific area, the reaction surface area, the rate $r_V$ and the local degree of conversion $x$. The latter is calculated from

$$x = 1 - \frac{1}{1-\varepsilon_0}\exp\left(-\int_0^\infty \pi R_r^2\, l(R_0)\,dR_0\right) \tag{18}$$

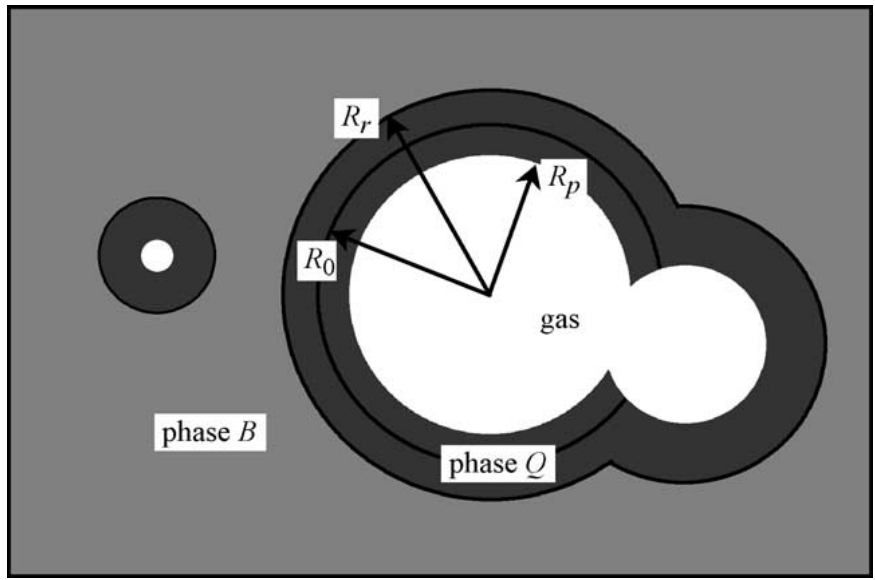

Fig. 2. Schematic representation of pore evolution.

## 2.4. Numerical method

The equations are discretized and solved using an implicit formulation of the finite volume method. The time step is adaptive and varies with the degree of conversion. A specific treatment of the thermal source term has been developed to simulate exothermic reactions with high activation energies [1].

## 3. Application

*3.1. Oxidation of zinc sulphide*

After validation of the numerical part of the model by comparison with different analytical and numerical solutions published in the literature [1], it was used to simulate a highly exothermic gas-solid reaction, the oxidation of zinc sulphide

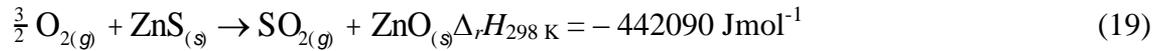

$$\tfrac{3}{2} O_{2(g)} + ZnS_{(s)} \rightarrow SO_{2(g)} + ZnO_{(s)} \quad \Delta_r H_{298\,K} = -442090 \text{ Jmol}^{-1} \qquad (19)$$

With the aid of the model, it was possible to quantitatively interpret experimental results obtained on highly reactive spherical pellets of synthetic ZnS. Pores in those pellets were large and did not change during conversion. No pore sub-model was then used. The diffusion coefficients $D_{ie}$ were calculated from the three binary diffisivities for a $N_2$-$O_2$-$SO_2$ mixture, without Knudsen diffusion.

We have shown [9] that, (i) because of the temperature rise due to the reaction, the kinetic regime was always diffusional, even when the external temperature conditions suggested a chemical or mixed regime, (ii) that the external resistance to matter transfer was not negligible, even for high gas flowrates, and (iii) that steep temperature gradients existed in the reacting pellet, localized in the peripheral oxide layer.

For one of the experiments, figure 3 compares the measured temperatures and degrees of conversion with those calculated using the model, as a function of time. The temperature is measured at the centre of the pellet (r = 0), and calculated at the centre (r = 0), at r = 3 mm and at the outer surface (r = 5 mm). The agreement between the measured and calculated values is quite satisfactory and provides a new validation of the model, this time experimental.

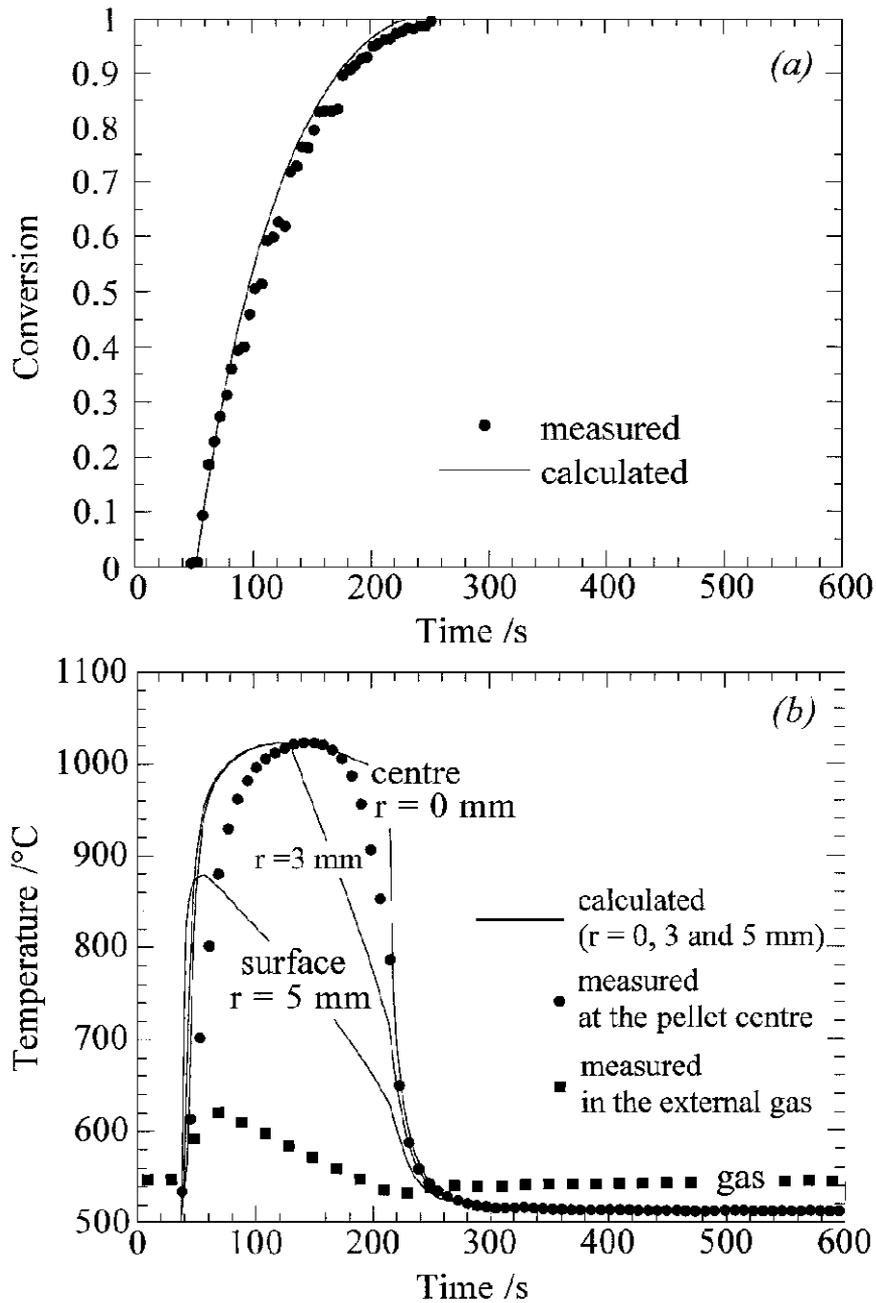

Fig. 3. Simulation of the oxidation of a zinc sulphide pellet. (a) Conversion vs. time. (b) Temperature vs. time. Furnace temperature: 550 j C; atmosphere: pure oxygen.

3.2. Hydrofluorination of uranium dioxide

The uranium dioxide hydrofluorination reaction

$$4\,HF_{(g)} + UO_{2(s)} \rightarrow 2\,H_2O_{(g)} + UF_{4(s)} \tag{20}$$

is one of the numerous gas-solid reactions involved in the manufacture of nuclear fuel. This reaction, which is exothermic and which involves a high volume expansion (the ratio $Z$ between the molar volumes of $UF_4$ and $UO_2$ is 1.88), is perfectly adapted to the version of our model describing the variation of the porosity.

Reaction rate measurements [10] have revealed a difference in reactivity between two dioxides, both composed of pure $UO_2$ pellets, but produced by different preparation processes. For example, the solid lines in figure 4 show the variation of the degree of conversion with time for the two dioxides, designated $D$ and $T$, hydrofluorinated in the same conditions (350 °C, $N_2$ + 25% HF). The dioxide $D$ is much more reactive than the dioxide $T$.

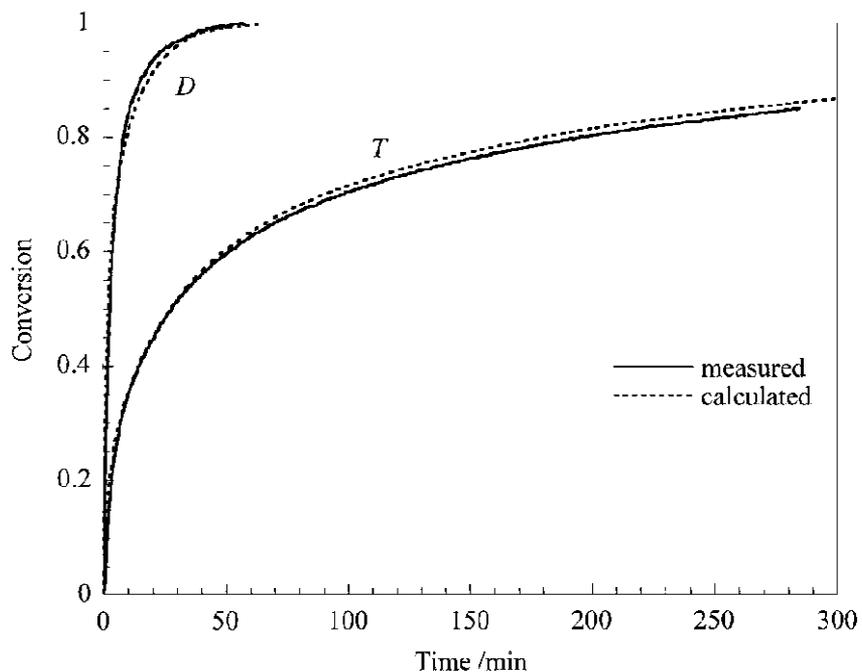

Fig. 4. Measured and calculated conversion versus time for two different uranium dioxides, D and T.

The simulation of these experiments using the model is represented by the dotted lines in figure 4. An excellent agreement can be seen between the measurements and calculations, due in particular to the fact that the unknown kinetic parameters (the rate constant $k_r$ and the diffusion coefficient in the solid product $D_{A(Q)}$) were determined by adjustment. However, the important point is that the same values of these kinetic parameters were used for both of the dioxides; only the physical parameters (apparent density, pellet diameter, and particularly the pore size distribution function) were different. The functions $l(R_0)$ employed were directly determined from the mercury porometry measurements for the two specimens given in figure 5, according to

$$I(R_0) = -\frac{1}{p R_0^2} \frac{1}{1-e_{>R_0}} \frac{de_{>R_0}}{dR_0} \tag{21}$$

where $e_{>R_0}$ represents the contribution to the porosity of pores with an initial radius greater than $R_0$. For calculating the gaseous diffusivities, molecular and Knudsen diffusion in a binary mixture was considered.

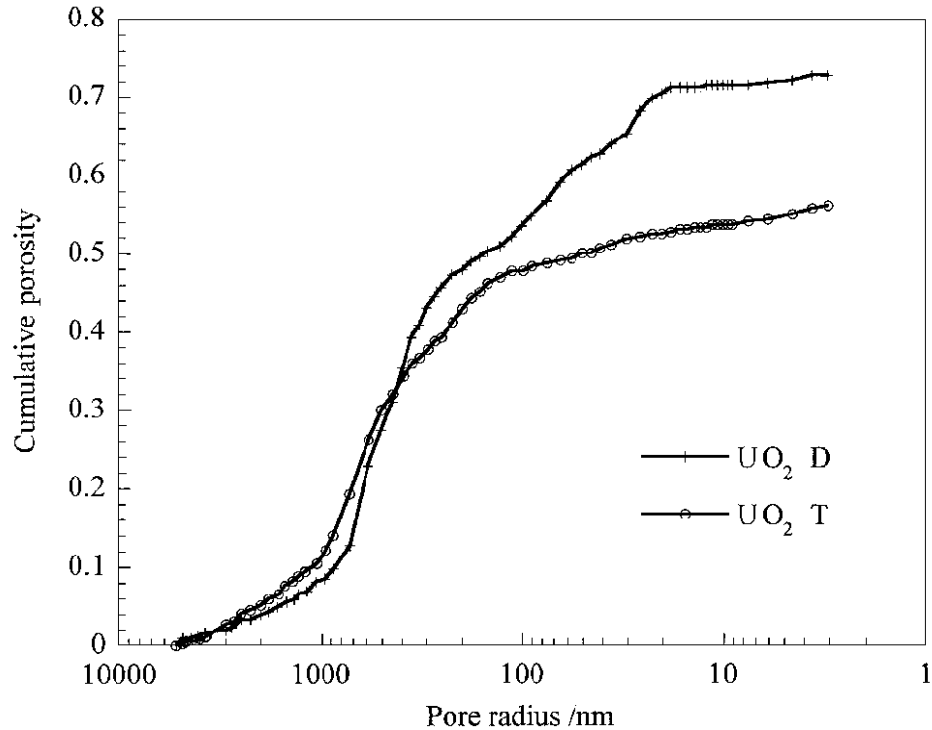

Fig. 5. Mercury intrusion curves for two uranium dioxide samples.

Figure 6 shows the variation of the porous structure between $UO_2$ and $UF_4$ for the dioxide *D* and provides a detailed understanding of how the reaction proceeds as a function of pore size. During hydrofluorination, the expansion of the solid volume due to the formation of $UF_4$ causes the pores to gradually shrink. The smallest ones (less than 60 nm in radius) are rapidly closed. The reaction continues around the largest pores, but at a slower rate, since they have a smaller specific surface area. For the dioxide *T*, which initially contained a lower fraction of fine pores (cf. figure 5), their closure is rapidly completed and the reaction occurs very slowly, as shown in figure 4. This result illustrates the ability of the model to describe differences in apparent conversion rates that can be entirely attributed in reality to differences in the porous structure.

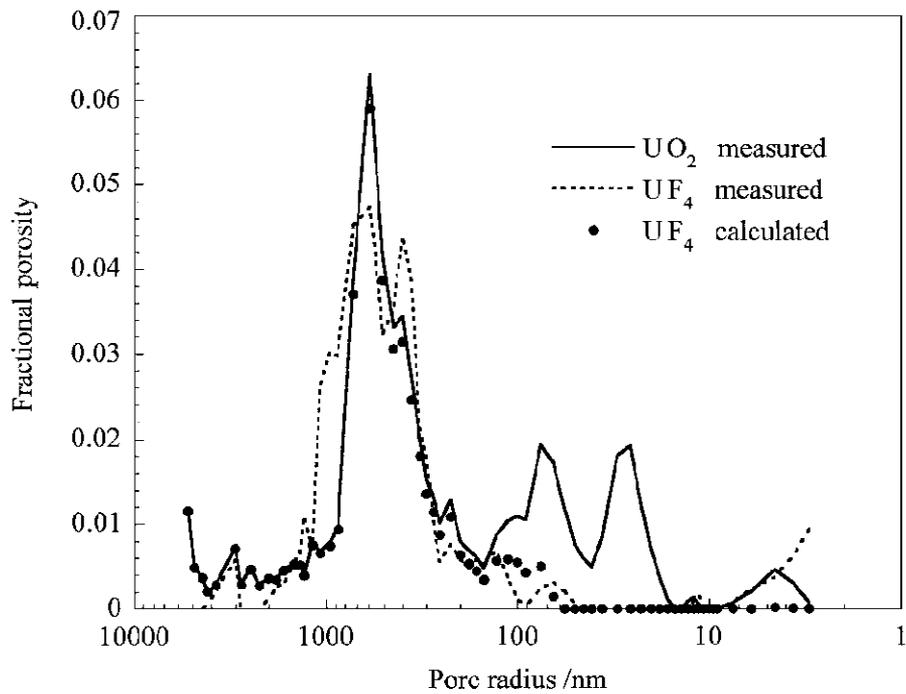

Fig. 6. Pore distribution of D uranium dioxide and tetrafluoride.

## 4. Conclusions

A kinetic model has been developed for gas-porous solid reactions that takes into account external matter transfer, internal transport, both by diffusion in pores of different sizes and by diffusion in the layer of solid reaction product, the chemical reaction itself, and the evolution or consumption of heat by the reaction and its transport by effective conduction. A pore sub-model is used to mathematically represent the complex structure of the porous solid by means of a unique function of the distribution of pore lengths with their diameter. This function can be readily determined from the results of mercury porometry experiments.

The model is capable of simulating numerous gas-solid reaction systems. It has been successfully applied to the treatment of a situation where the thermal effects associated with the reaction are decisive (the exothermic oxidation of zinc sulphide pellets) and also to a case where the variation of the porous structure of the solid controls its reactivity (the hydrofluorination of uranium dioxide).

**Acknowledgements**

The authors are extremely grateful to Mr. F. Nicolas, Research Manager at the Comurhex plant in Pierrelatte, for his participation in the uranium dioxide hydrofluorination study.

**References**

[1]   F. Patisson, M. Galant François, D. Ablitzer, Chem. Eng. Sci. 53 (1998) 697-708.

[2]   J. Szekely, J.W. Evans, H.Y. Sohn, Gas-solid reactions, Academic Press, New York, 1976.

[3]   A.B.M. Heesink, W. Prins, W.P.M. van Swaaij, Chem. Eng. J. 53 (1993) 25-37.

[4]   A.B.M. Heesink, D.W.F. Brilman, W.P.M. van Swaaij, AIChE J. 44 (1998) 1657-1666.

[5]   P.A. Ramachandran, J.M. Smith, AIChE J. 23 (1977) 353-361.

[6]   S.K. Bhatia, AIChE J. 31 (1985) 642-648.

[7]   S.V. Sotirchos, S. Zarkanitis, Chem. Eng. Sci. 48 (1993) 1487-1502.

[8]   S. Yagi, D. Kunii, Proceedings of 5th (Int.) Symposium on Combustion, Reinhold, New York, 1955, 231-244.

[9]   M. Galant François, Etude cinétique de la réaction exothermique d'oxydation du sulfure de zinc, Doctorate thesis, Institut National Polytechnique de Lorraine, Nancy, 1995.

[10] F. Sbaffo, Cinétique et modélisation d'une réaction gaz-solide avec évolution de porosité: l'hydrofluoration du dioxyde d'uranium, DEA dissertation, INPL, Nancy, 1998.